\documentclass[%
reprint,
superscriptaddress,
 amsmath,amssymb,
 aps,
 prx,
floatfix,
a4paper
]{revtex4-2}

\usepackage[colorlinks=true,allcolors=blue]{hyperref}

\usepackage{graphicx}%
\usepackage{hyperref}
\usepackage{polyglossia}
\setdefaultlanguage{english}
\begin{document} 

\title{Attosecond soft X-ray pulses generated by chirp-dispersed manipulation in an XFEL reveal nonlinear core-electron dynamics in neon}

% RevTeX 4.2 author block, converted from sn-jnl

\author{Lars Funke}
\email{lars.funke@tu-dortmund.de}
\thanks{L.~Funke, M.~Ilchen, and S.~Serkez contributed equally to this work.}
\affiliation{Fakultät Physik, Technische Universität Dortmund, Dortmund, Germany}

\author{Markus Ilchen}
\email{markus.ilchen@uni-hamburg.de}
\affiliation{Institut für Experimentalphysik, Universität Hamburg, Hamburg, Germany}
\affiliation{European X-Ray Free-Electron Laser Facility GmbH, Schenefeld, Germany}
\affiliation{Deutsches Elektronen-Synchrotron DESY, Hamburg, Germany}

\author{Svitozar Serkez}
\email{svitozar.serkez@xfel.eu}
\affiliation{European X-Ray Free-Electron Laser Facility GmbH, Schenefeld, Germany}

\author{Kristina Scharei}
\affiliation{Intelligent Embedded Systems, Universität Kassel, Kassel, Germany}
\affiliation{Artificial Intelligence Methods for Experiment Design (AIM-ED), Joint Lab Helmholtzzentrum für Materialien und Energie, Berlin (HZB) and University of Kassel, Berlin, Germany}

\author{Tommaso Mazza}
\affiliation{European X-Ray Free-Electron Laser Facility GmbH, Schenefeld, Germany}

\author{Terence Mullins}
\affiliation{European X-Ray Free-Electron Laser Facility GmbH, Schenefeld, Germany}
\affiliation{Deutsches Elektronen-Synchrotron DESY, Hamburg, Germany}
\affiliation{The Hamburg Centre for Ultrafast Imaging, Universität Hamburg, Hamburg, Germany}

\author{Thorsten Otto}
\affiliation{Deutsches Elektronen-Synchrotron DESY, Hamburg, Germany}

\author{Daniel E. Rivas}
\affiliation{European X-Ray Free-Electron Laser Facility GmbH, Schenefeld, Germany}

\author{Sara Savio}
\affiliation{Fakultät Physik, Technische Universität Dortmund, Dortmund, Germany}

\author{Peter Walter}
\affiliation{Institut für Experimentalphysik, Universität Hamburg, Hamburg, Germany}
\affiliation{SLAC National Accelerator Laboratory, Menlo Park, CA, USA}
\affiliation{TAU Systems, Austin, TX, USA}

\author{Niclas Wieland}
\affiliation{Institut für Experimentalphysik, Universität Hamburg, Hamburg, Germany}

\author{Lasse Wülfing}
\affiliation{Fakultät Physik, Technische Universität Dortmund, Dortmund, Germany}

\author{Sadia Bari}
\affiliation{Deutsches Elektronen-Synchrotron DESY, Hamburg, Germany}
\affiliation{Zernike Institute for Advanced Materials, University of Groningen, Groningen, The Netherlands}

\author{Rebecca Boll}
\affiliation{European X-Ray Free-Electron Laser Facility GmbH, Schenefeld, Germany}

\author{Markus Braune}
\affiliation{Deutsches Elektronen-Synchrotron DESY, Hamburg, Germany}

\author{Francesca Calegari}
\affiliation{Deutsches Elektronen-Synchrotron DESY, Hamburg, Germany}

\author{Alberto De Fanis}
\affiliation{European X-Ray Free-Electron Laser Facility GmbH, Schenefeld, Germany}

\author{Winfried Decking}
\affiliation{Deutsches Elektronen-Synchrotron DESY, Hamburg, Germany}

\author{Andreas Duensing}
\affiliation{Physik-Department, TUM School of Natural Sciences, Technische Universität München, Garching, Germany}

\author{Stefan Düsterer}
\affiliation{Deutsches Elektronen-Synchrotron DESY, Hamburg, Germany}

\author{Felix Egun}
\affiliation{Blackett Laboratory, Imperial College London, London, United Kingdom}
\affiliation{Center for Free-Electron Laser Science, Hamburg, Germany}

\author{Arno Ehresmann}
\affiliation{Institut für Physik und CINSaT, Universität Kassel, Kassel, Germany}
\affiliation{Artificial Intelligence Methods for Experiment Design (AIM-ED), Joint Lab Helmholtzzentrum für Materialien und Energie, Berlin (HZB) and University of Kassel, Berlin, Germany}

\author{Benjamin Erk}
\affiliation{Deutsches Elektronen-Synchrotron DESY, Hamburg, Germany}

\author{Danilo Enoque Ferreira de Lima}
\affiliation{European X-Ray Free-Electron Laser Facility GmbH, Schenefeld, Germany}

\author{Andreas Galler}
\affiliation{European X-Ray Free-Electron Laser Facility GmbH, Schenefeld, Germany}

\author{Gianluca Geloni}
\affiliation{European X-Ray Free-Electron Laser Facility GmbH, Schenefeld, Germany}

\author{Natalia Gerasimova} %added
\affiliation{European X-Ray Free-Electron Laser Facility GmbH, Schenefeld, Germany}

\author{Marc Guetg}
\affiliation{Deutsches Elektronen-Synchrotron DESY, Hamburg, Germany}

\author{Jan Grünert}
\affiliation{European X-Ray Free-Electron Laser Facility GmbH, Schenefeld, Germany}

\author{Patrik Grychtol}
\affiliation{European X-Ray Free-Electron Laser Facility GmbH, Schenefeld, Germany}

\author{Andreas Hans}
\affiliation{Institut für Physik und CINSaT, Universität Kassel, Kassel, Germany}
\affiliation{Artificial Intelligence Methods for Experiment Design (AIM-ED), Joint Lab Helmholtzzentrum für Materialien und Energie, Berlin (HZB) and University of Kassel, Berlin, Germany}

\author{Arne Held}
\affiliation{Fakultät Physik, Technische Universität Dortmund, Dortmund, Germany}

\author{Ruda Hindrikson}
\affiliation{Intelligent Embedded Systems, Universität Kassel, Kassel, Germany}

\author{Till Jahnke}
\affiliation{Max-Planck-Institut für Kernphysik, Heidelberg, Germany}

\author{Joakim Laksman}
\affiliation{European X-Ray Free-Electron Laser Facility GmbH, Schenefeld, Germany}

\author{Mats Larsson}
\affiliation{Department of Physics, AlbaNova University Center, Stockholm University, Stockholm, Sweden}

\author{Jia Liu}
\affiliation{European X-Ray Free-Electron Laser Facility GmbH, Schenefeld, Germany}

\author{Jon P. Marangos}
\affiliation{Blackett Laboratory, Imperial College London, London, United Kingdom}

\author{Lutz Marder}
\affiliation{Institut für Physik und CINSaT, Universität Kassel, Kassel, Germany}
\affiliation{Artificial Intelligence Methods for Experiment Design (AIM-ED), Joint Lab Helmholtzzentrum für Materialien und Energie, Berlin (HZB) and University of Kassel, Berlin, Germany}

\author{David Meier}
\affiliation{Optik und Strahlrohre, Helmholtz-Zentrum Berlin für Materialien und Energie GmbH, Berlin, Germany}
\affiliation{Artificial Intelligence Methods for Experiment Design (AIM-ED), Joint Lab Helmholtzzentrum für Materialien und Energie, Berlin (HZB) and University of Kassel, Berlin, Germany}

\author{Michael Meyer}
\affiliation{European X-Ray Free-Electron Laser Facility GmbH, Schenefeld, Germany}

\author{Najmeh Mirian}
\affiliation{Helmholtz-Zentrum Dresden-Rossendorf, Dresden, Germany}

\author{Christian Ott}
\affiliation{Max-Planck-Institut für Kernphysik, Heidelberg, Germany}

\author{Christopher Passow}
\affiliation{Deutsches Elektronen-Synchrotron DESY, Hamburg, Germany}

\author{Thomas Pfeifer}
\affiliation{Max-Planck-Institut für Kernphysik, Heidelberg, Germany}

\author{Patrick Rupprecht}
\affiliation{Max-Planck-Institut für Kernphysik, Heidelberg, Germany}
\affiliation{Chemical Sciences Division, Lawrence Berkeley National Laboratory, Berkeley, CA, USA}
\affiliation{Department of Chemistry, University of California, Berkeley, CA, USA}
\affiliation{Albert-Ludwigs-Universität Freiburg, Freiburg, Germany}

\author{Albert Schletter}
\affiliation{Physik-Department, TUM School of Natural Sciences, Technische Universität München, Garching, Germany}

\author{Philipp Schmidt}
\affiliation{European X-Ray Free-Electron Laser Facility GmbH, Schenefeld, Germany}

\author{Frank Scholz}
\affiliation{Deutsches Elektronen-Synchrotron DESY, Hamburg, Germany}

\author{Simon Schott}
\affiliation{Intelligent Embedded Systems, Universität Kassel, Kassel, Germany}

\author{Evgeny Schneidmiller}
\affiliation{Deutsches Elektronen-Synchrotron DESY, Hamburg, Germany}

\author{Bernhard Sick}
\affiliation{Intelligent Embedded Systems, Universität Kassel, Kassel, Germany}
\affiliation{Artificial Intelligence Methods for Experiment Design (AIM-ED), Joint Lab Helmholtzzentrum für Materialien und Energie, Berlin (HZB) and University of Kassel, Berlin, Germany}

\author{Kai Tiedtke}
\affiliation{Deutsches Elektronen-Synchrotron DESY, Hamburg, Germany}

\author{Sergey Tomin} %added
\affiliation{Deutsches Elektronen-Synchrotron DESY, Hamburg, Germany}

\author{Andrei Trebushinin} %added
\affiliation{European X-Ray Free-Electron Laser Facility GmbH, Schenefeld, Germany}

\author{Sergey Usenko}
\affiliation{European X-Ray Free-Electron Laser Facility GmbH, Schenefeld, Germany}

\author{Vincent Wanie}
\affiliation{Deutsches Elektronen-Synchrotron DESY, Hamburg, Germany}

\author{Markus Wurzer}
\affiliation{Physik-Department, TUM School of Natural Sciences, Technische Universität München, Garching, Germany}

\author{Mikhail Yurkov}
\affiliation{Deutsches Elektronen-Synchrotron DESY, Hamburg, Germany}

\author{Igor Zagorodnov} %added
\affiliation{Deutsches Elektronen-Synchrotron DESY, Hamburg, Germany}

\author{Vitali Zhaunerchyk}
\affiliation{Department of Physics, University of Gothenburg, Gothenburg, Sweden}

\author{Wolfram Helml}
\email{wolfram.helml@tu-dortmund.de}
\affiliation{Fakultät Physik, Technische Universität Dortmund, Dortmund, Germany}

\date{\today}

\begin{abstract}
Free-electron lasers deliver intense attosecond X-ray pulses, which are the key to studying electron dynamics in specifically targeted electronic states, but come at the expense of complicated generation schemes and stochastic pulse shapes.
Here, we demonstrate a novel and simple operation concept based on dispersive manipulation of the lasing electron bunch and capable of working with the full 4.5 MHz repetition rate at the European XFEL in Germany. Using a high-fidelity single-shot temporal characterisation, we detect X-ray pulses with durations of down to 200 attoseconds and peak powers reaching into the terawatt regime at $\sim$1\,keV photon energy.
As a direct application, we present simultaneous measurements of nonlinear X-ray--matter interaction via time-resolved electron spectroscopy. Using the pulse duration retrieval and restricting  the durations to a regime where individual X-ray pulses are shorter than the single-core-hole life time in neon atoms, we reveal an otherwise hidden peak-intensity dependence in the nonlinear dynamics of double-core-hole formation. Our results shed light on highly transient intermediate steps in complex electronic dynamics of systems driven far off their equilibrium state, and thus promise to help build the conceptual bridge between fundamental physics processes and real-world photo reactions.
\end{abstract}

\maketitle
%\linenumbers

\section*{Introduction}
\label{sec:Intro}

%Allgemeine Einführung
Attosecond physics has been developed from an experimental novelty to a Nobel prize-winning area of fundamental research~\cite{paul2001observation, hentschel2001attosecond}. It has been extended to measurements of electron tunnelling in atoms~\cite{uiberacker2007attosecond}, timing the photoemission from metal surfaces and bulk material~\cite{cavalieri2007attosecond} as well as investigating excited state dynamics in atoms~\cite{Goulielmakis2010}, molecules~\cite{Sansone2010, Kraus2015} and in the formation of plasmons~\cite{stockman2007attosecond}. Nevertheless, many desirable measurements have been hampered by the limited energy tunability and relatively low flux at higher photon energies than the extreme ultraviolet range achievable with high harmonic generation (HHG) sources~\cite{lewenstein1994theory, shiner2009wavelength}.

X-ray free-electron lasers (XFELs) exhibit the potential to provide both tunability and high intensity of sub-femtosecond pulses in a wide range of photon energies from soft to hard X-rays~\cite{marinelli2017experimental,malyzhenkov2020,prat2023coherent}.
The phase space of the highly relativistic electron bunches in the FEL process can be tailored to ensure lasing with attosecond pulse durations \cite{zholents2004,Emma2004,saldin2004,xiang2009,tanaka2013,prat2015,wang2017,Huang2017,lutman2016fresh,lutman2018,serkez2018overview,duris2020tunable,zhang2020experimental,Li2024_beam,robbles2025,yan2024terawatt}, as demonstrated in pioneering work showing attosecond X-ray time--energy structures \cite{hartmann2018attosecond} and studying coherent electron motion in molecules~\cite{li2022attosecond}.
Particularly noteworthy are the latest achievements at FELs showcasing attosecond pump/probe capabilities for measurements in the gaseous~\cite{Guo2024} and the liquid phase~\cite{Li2024}, as well as studies of nonlinear X-ray interactions in impulsive stimulated Raman scattering with attosecond pulses in gas phase molecules \cite{ONealPRL2020} and in liquid water \cite{ScienceAdvAlexander2024}.

Crucial details of light--matter interactions, such as coherent electron motion in nonlinearly populated states~\cite{young2018roadmap,Eichmann2020,Alexander2024}, fundamentally depend on the exact power profiles of X-ray attosecond pulse structures, varying for successive shots due to the stochastic nature of self-amplified spontaneous emission (SASE)~\cite{Milton2001}. Several attempts to predict temporal pulse characteristics using standard accelerator electron and photon diagnostics alongside machine-learning (ML) approaches~\cite{sanchez2017, AlaaEl-Din2024} have been made. Accurate retrieval relying solely on these indirect measurements, let alone direct control of SASE FEL time--energy structures has not been achieved, especially with highly manipulated electron bunches for the production of attosecond pulses.
High-repetition-rate XFELs \cite{decking2020mhz} can help mitigate these limitations by using the intrinsic SASE variability of the pulse structure distribution in combination with a non-invasive pulse-by-pulse diagnostic and a near-online spectro-temporal analysis method based on 'angular streaking' \cite{hartmann2018attosecond, Li2018a, heider2019megahertz}.

While the aforementioned methods usually involve the installation of additional hardware, we present a novel approach to reduce pulse durations, involving the manipulation of both the longitudinal and transverse phase spaces of the electron beam, exploiting the existing infrastructure of a typical X-ray FEL facility. In this scheme, the generated set of stochastic pulse shapes can be sorted for individual durations. In recent experiments, angular streaking has already been used to determine an overall regime of attosecond pulse durations for specific FEL operation modes, without considering the specific single-shot X-ray pulse duration as a parameter for further analyses~\cite{li2022attosecond, Guo2024, Li2024}.
We have overcome this methodological barrier via simultaneous measurement of the electron reconfiguration dynamics in a core-ionised neon target with parallel full characterisation of individual SASE pulses.
%This is not merely an incremental step towards advanced temporal resolution, but rather the opening of a new field of attosecond-resolved exploration with XFELs.

To define a common and reproducible measure for the SASE pulse shapes described in this work, in the following we consistently use the term \emph{pulse duration (FWHM)}, which is given as the time span between the first and last occurrence of half the maximum intensity within a single shot. Note that, due to the stochastic nature of the pulses, crossing the half maximum intensity value may occur just twice (for a single spike) or multiple times (for few-spike pulses) per shot.% Throughout the paper, we refer to this measure as the (FWHM) pulse duration.
\section*{Generation of attosecond X-ray FEL pulses}
\label{sec:Attopulses_exp}

\onecolumngrid

\begin{figure}
    \centering
    \includegraphics[width=1\textwidth]{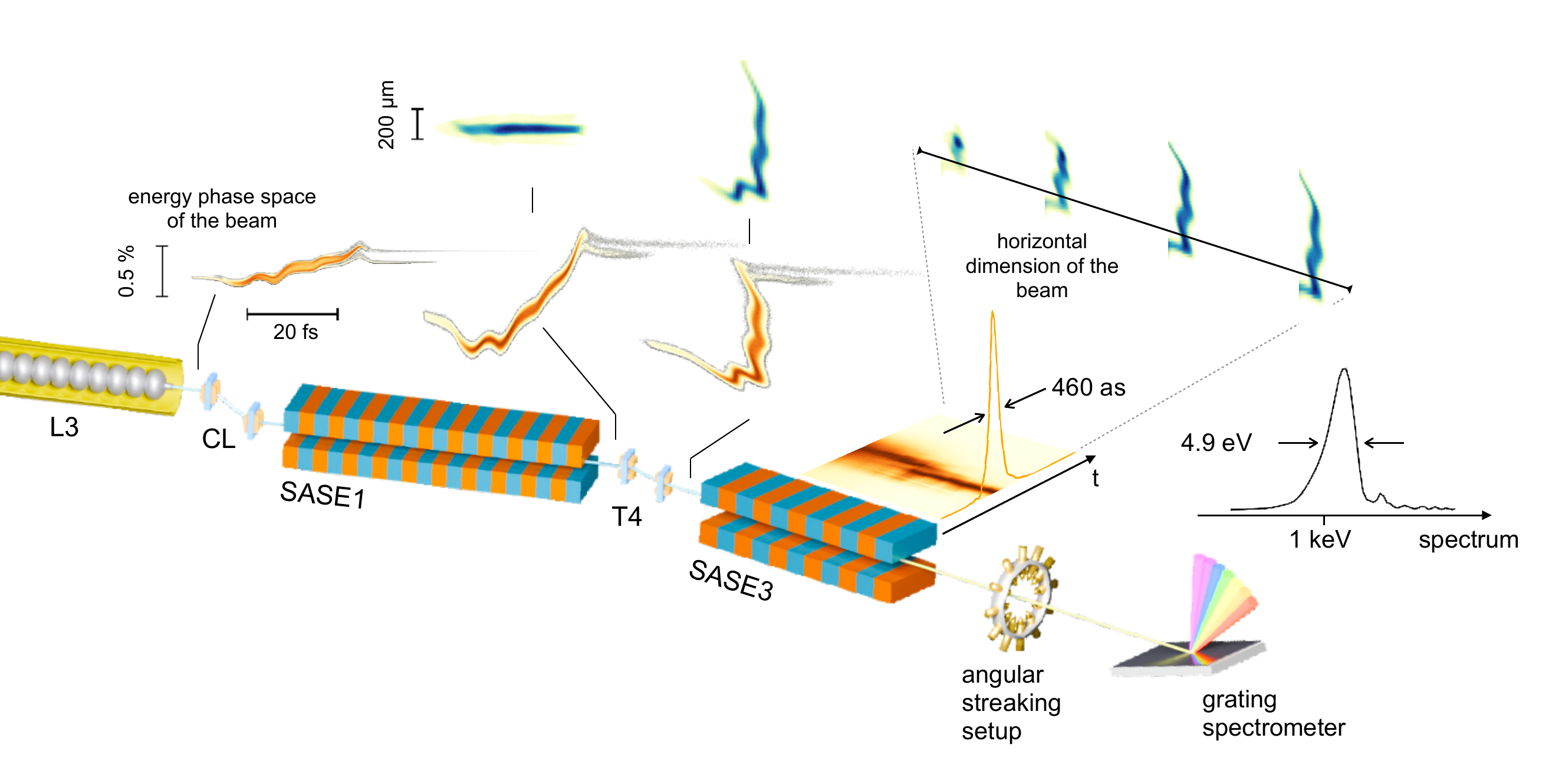}
    \caption{Electron beam compression scheme to generate attosecond pulses at the European XFEL. The linearly chirped electron beam is strongly compressed upstream of the L3 part of the accelerator and its dispersion evolves in a controlled way as it is transported through the following sections (labelled CL \& SASE1). In an arc before the SASE3 undulator (T4),  the energy-chirped electron beam is compressed further and acquires a horizontal tilt. We provide simulations of the evolution of its energy chirp (red-yellow colormap) and its transverse size (blue-green colormap) as well as the development of the normalised radiation power along the electron beam as it propagates through the SASE3 undulator (flat image next to SAS3). During the experiment, the FEL temporal properties are directly diagnosed with angular streaking, as illustrated in \autoref{fig:exp} and indirectly with a plane grating spectrometer (not simultaneously).}
    \label{fig:placeholder}
\end{figure}
\twocolumngrid

At the SASE3 undulator line of the Europe   an XFEL, we demonstrate generation of high-power, soft X-ray attosecond SASE pulses at MHz repetition rate by increasing the FEL gain and shortening the effective lasing window. We accomplish this by creating an energy chirp upon the electron beam acceleration and transport, and then exploiting this chirp in the undulator through non-standard control of the transverse and longitudinal dispersion (see Figure~\ref{fig:placeholder}).% We refer to this method as the chirp--dispersion method or CHIPS.

% This can be moved or re-written/toned down in the "Generation" paragraph
% A quadrupole magnet creates a highly nonlinear correlation between the energies, momenta, and positions of the electrons in the bunch which is used to dramatically increase the charge density and narrow the effective lasing window. 
% This manipulation extends to the full six-dimensional phase space.
% This process is naturally assisted and enhanced by the self-interaction of charges, also known as space-charge effects.
% Furthermore, the angular dispersion of radiation enables single attosecond spikes to be spatially filtered, thereby eliminating potential background radiation from less dense regions of the beam.

As a first step of tuning the machine from the nominal SASE regime to deliver attosecond pulses, we maximise the energy chirp along the electron bunch. To achieve this, we optimise the compression setup such that, with the additional aid of resistive wakefields, upon exiting the first section of the machine in front of the SASE3 undulator the electron energy drops by about 1~\% from the head towards the tail of the bunch, as shown in the according inset of Figure~\ref{fig:placeholder}, downstream of the SASE1 undulator.

This chirped beam passes through a double-bend achromat arc (T4 in Fig.~\ref{fig:placeholder}) that hosts two doublets of quadrupoles in its dispersive section. By adjusting the magnetic field in one of the quadrupole doublets, we simultaneously control both transverse~\cite{guetg2018dispersion} and longitudinal \cite{yan2024terawatt} dispersion functions in the beam transport before entering the SASE3 soft X-ray undulator. In Figure~\ref{fig:placeholder}, downstream of the T4 arc, one can observe the compressed and tilted electron beam. With such tuning, we introduce 15~cm of transverse horizontal dispersion and 0.7~mm of positive longitudinal dispersion, resulting in both compression of the negatively chirped electron beam to about 50~kA peak current and its transverse tilt such that the projected horizontal size of the lasing window reaches an order of 1~mm. We depict this simulated electron beam after the T4 arc and its subsequent evolution in Figure~\ref{fig:placeholder} with orange (longitudinal phase space) and blue (horizontal projection) distribution colormaps.

% In this beam time, the intra-train repetition rate was 376\,kHz, but the generation scheme was also demonstrated at full repetition rate of 4.5\,MHz as well.
% The following information seems too specific here, but should be definitely given in the SI.
% We introduce 15~cm of transverse horizontal dispersion and 0.7~mm of positive longitudinal dispersion, resulting in both compression of the negatively chirped electron beam to about 50~kA peak current and its transverse tilt such that the projected horizontal size of the lasing window reaches an order of 1~mm.

In this way, the electron beam effectively undergoes transverse rotation and travels with a tilt through the undulator. Only a small slice of it is moving in a straight trajectory, being unperturbed by the T4 achromatic magnet lattice. In addition, space-charge effects dynamically increase the energy chirp in the undulator during the FEL amplification process. The resulting angularly dispersed femtosecond-order pulses were additionally shortened by geometrical filtering in the beamline apertures. As the method does not involve any external lasers or special synchronization, it is applicable and has been implemented at the maximum repetition rate of 4.5~MHz at the European XFEL, while in the current beam time the intra-train repetition rate was set to 376\,kHz.

The described method for the experimental generation of attosecond pulses was replicated with simulations, yielding comparable radiation properties as in the measurements. We combined the program ASTRA~\cite{Astra} with the Ocelot simulation package~\cite{Agapov2014a} and Genesis FEL code~\cite{GENE} version 4, for start-to-end simulation of electron beam formation, its transport through the accelerator, generation of attosecond pulses and their propagation, as illustrated in the insets of Figure~\ref{fig:placeholder}.

In the simulation we specifically accounted for the longitudinal space charge during the FEL amplification process.
Replication of accelerator parameters within diagnostics tolerances allowed us to follow the electron bunch shaping throughout the accelerator stages and to reproduce sub-femtosecond single-peak and double-peak FEL radiation, an example of which is illustrated in Figure~\ref{fig:IAS} \textbf{e)}.

\section*{SASE pulse single-shot characterisation}
\label{sec:SASE}

\begin{figure*}[t]
\includegraphics[width=1\textwidth]{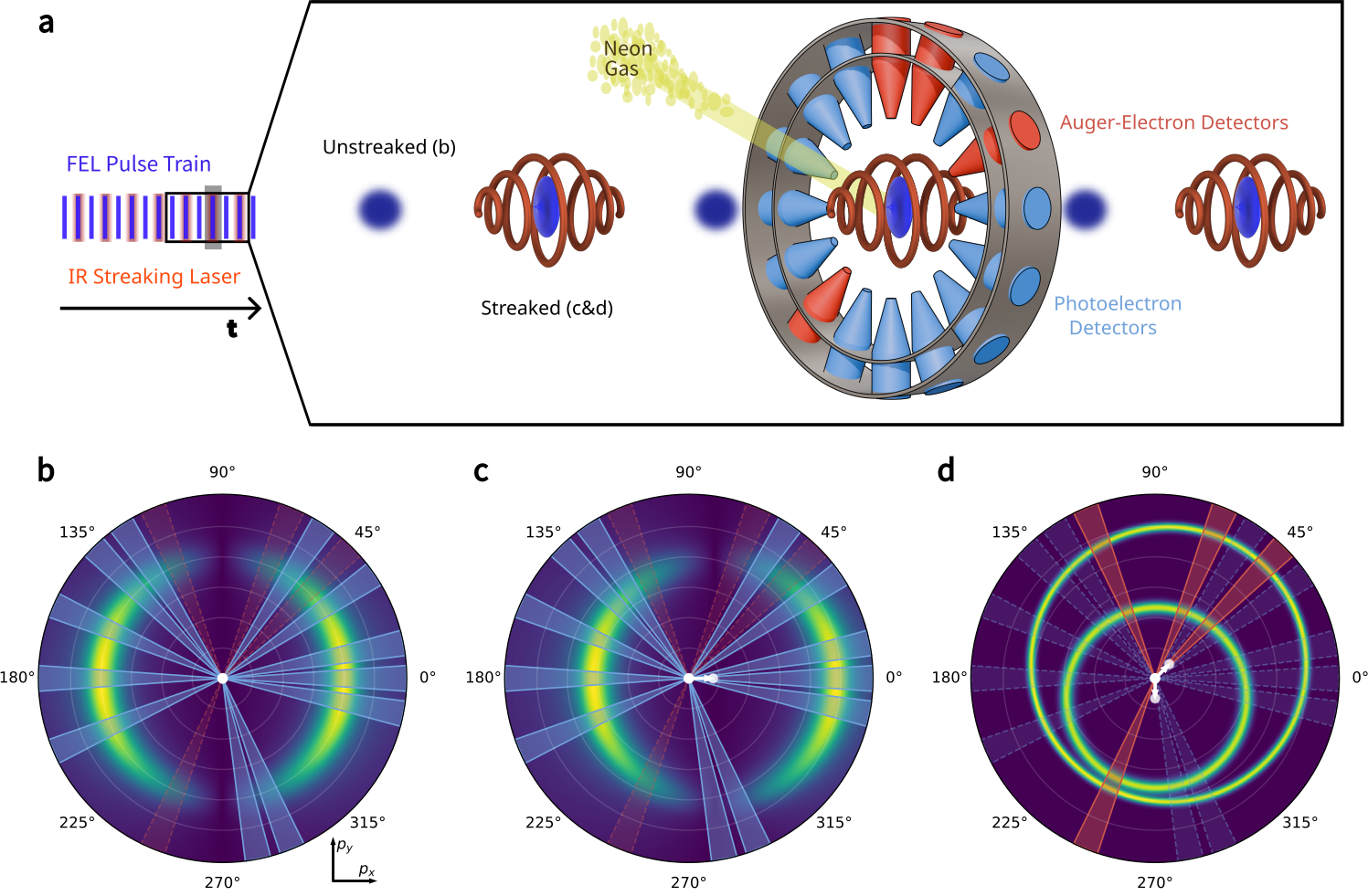}
\caption{\textbf{a)} Schematic overview of the angular streaking experiment. Every other in a train of XFEL pulses is overlapped with circularly polarised mid-infrared streaking laser pulses. The resulting MIR phase-dependent electron energy shift is measured in a time-of-flight photoelectron spectrometer array. The array is tuned to Auger electron energies for one set of four out of 16 detectors (red) and for 1s photoelectron energies for the other twelve detectors (blue).
\textbf{b)}--\textbf{d)} Three illustrative examples of simulated electron momentum spectra from neon in the polarisation plane ($\vec{p}_x \times \vec{p}_y$) of horizontally polarised attosecond XFEL pulses. The coloured wedges represent electron spectrometers at the actual angles used in the experiment. The momentum is zero in the origin (central white dot), and increasing radially outward. \textbf{b)} 1s photoelectrons in an unstreaked shot (no momentum shift), showing the typical anisotropic electron emission distribution of a spherical orbital. \textbf{c)} 1s photoelectrons from a single streaked XFEL pulse, shifting the \textit{photo}electron momentum distribution to the right (white arrow). \textbf{d)} Streaked spectra of single-core-hole and double-core-hole \textit{Auger} electrons, whose angular offsets reflect the time delay between the two processes like the hands of a clock, where one full rotation corresponds to the period of the MIR streaking laser. 
}
\label{fig:exp}
\end{figure*}

Experimentally, as the standard procedure the temporal characteristics of the pulses are indirectly inferred either from destructive, i.e. photon-absorbing, measurements of the X-ray pulse spectra via a grating spectrometer \cite{gerasimova2022soft} or from cross-correlation measurements \cite{rivas2022, grychtol2021}, the latter providing only average pulse durations.

For this experiment at the SQS (Small Quantum Systems) instrument of European XFEL, we employ angular streaking, as illustrated in \autoref{fig:exp}. In this scheme, a circularly polarised mid-infrared (MIR) laser pulse is temporally and spatially overlapped with the X-ray FEL pulse in a region flooded with neon gas at a partial pressure of approximately $10^{-7}$\,hPa. 1s-electrons of the neon atoms are ionised by the linearly polarised XFEL pulse. The outgoing photoelectrons interact with the circularly polarised 4.75\,µm-wavelength strong-field streaking pulse, changing their momenta and leading to an energy and angular redistribution of the electrons depending on the electric field amplitude and phase of the laser during the window of temporal overlap between X-ray and MIR pulses. The resulting angular and spectral distribution is detected by an array of 16 electron time-of-flight (eTOF) spectrometers, positioned in a plane perpendicular to the joint propagation direction of the MIR laser beam and the FEL around the interaction point ~\cite{hartmann2018attosecond}.

The spectral and temporal characteristics of individual pulses have been reconstructed using a post-experiment iterative retrieval code similar to the one published by Hartmann et al.~\cite{hartmann2018attosecond}, see also Sec. S2.2 in the SI. In addition, we tested a recently developed machine-learning technique~\cite{dingel2022artificial}, which allows for similar reconstruction quality using less computation time and bearing the potential for online pulse characterisation (Sec. S2.5 in the SI).

Angular streaking is the only currently available direct and non-invasive method for characterizing the temporal and spectral X-ray structures of single FEL pulses, i.e., the number of SASE peaks (and their sub-femtosecond durations), as well as the spectral chirp for each single shot. This knowledge can be used to find optimised FEL settings for targeted X-ray pulse characteristics, such as the shortest possible pulses, double-peak pulses with defined delay, or two-colour operation modes, as, for example, demonstrated in \cite{Li2024_beam}. The streaking technique in a low-density gas target exerts a negligible influence on the X-rays, with an absorption fraction on the sub-percent level. Consequently, these same pulses can be used in a parallel experiment, which can now capitalise on the fully determined X-ray characteristics via pulse tagging. 

\begin{figure}
\includegraphics[width=1\linewidth]{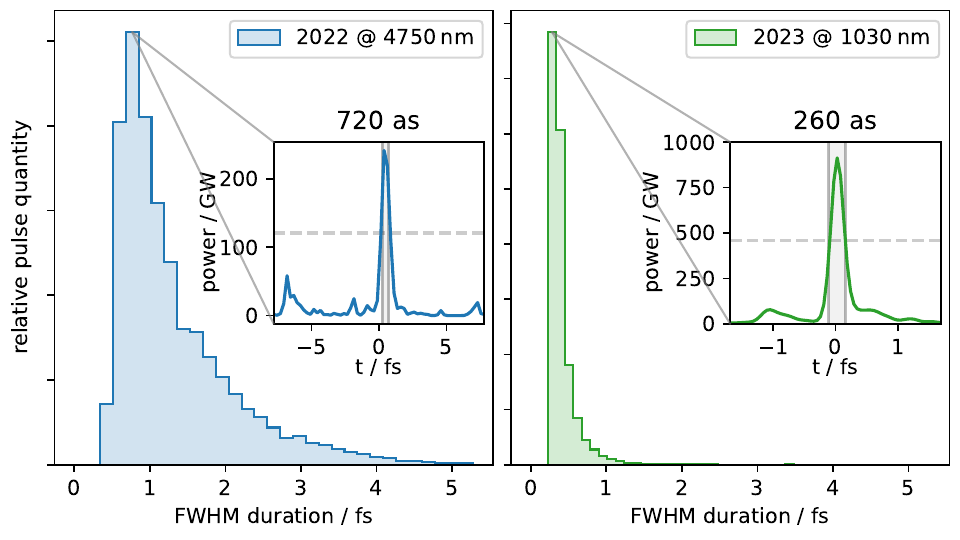} 
\caption{Distribution of pulse duration (FWHM) for two evaluated beam times, corresponding to two different streaking wavelengths. Left: Data from 2022 taken with a streaking wavelength of 4.75\,µm (73176 shots). Right: Data from 2023 taken with a streaking wavelength of 1030\,nm (100000 shots). Insets: Exemplary single pulse temporal shapes from the respective data sets with indicated FWHM measurements. The temporal resolution of the reconstruction is estimated as 0.3\,fs in 2022 and 0.22\,fs in 2023 (for more details see the SI).
}
\label{fig:modeoverview}
\end{figure}

\section*{Terawatt-scale attosecond X-ray pulses}
\label{sec:highAS}
In order to demonstrate the time-resolving capabilities, time--energy spectrograms for pulses with individual single-peak powers of hundreds of GW have been extracted. Knowledge about the pulse structure allows sorting on pulse duration as well as applying further analysis methods such as stochastic delay scans. As the European XFEL can deliver  $>$1000 shots per second, this 'post-sorting' yields the promise to get access to X-ray pump/X-ray probe methodology with highly intense pulses and attosecond time resolution without the need of additional infrastructure, like split-and-delay beamlines.
We identified isolated attosecond soft X-ray pulses at photon energies around 1\,keV, with a peak power of up to 200\,GW in a first beamtime in the year 2022 and up to one TW in a second beamtime in 2023.

In the first beamtime, approximately 38\,\% of the shots were found to be shorter than 1\,fs in this data set, and approx. 17\,\% below 750\,as, with the shortest pulses in the order of 600\,as (blue distribution on the left in \autoref{fig:modeoverview}). In the second beamtime, due to a more optimised electron bunch compression setting, over 99\% of the shots were found to be in the attoregime (green distribution on the right in \autoref{fig:modeoverview}).

\begin{figure}
    \centering
    \includegraphics[width=1\linewidth]{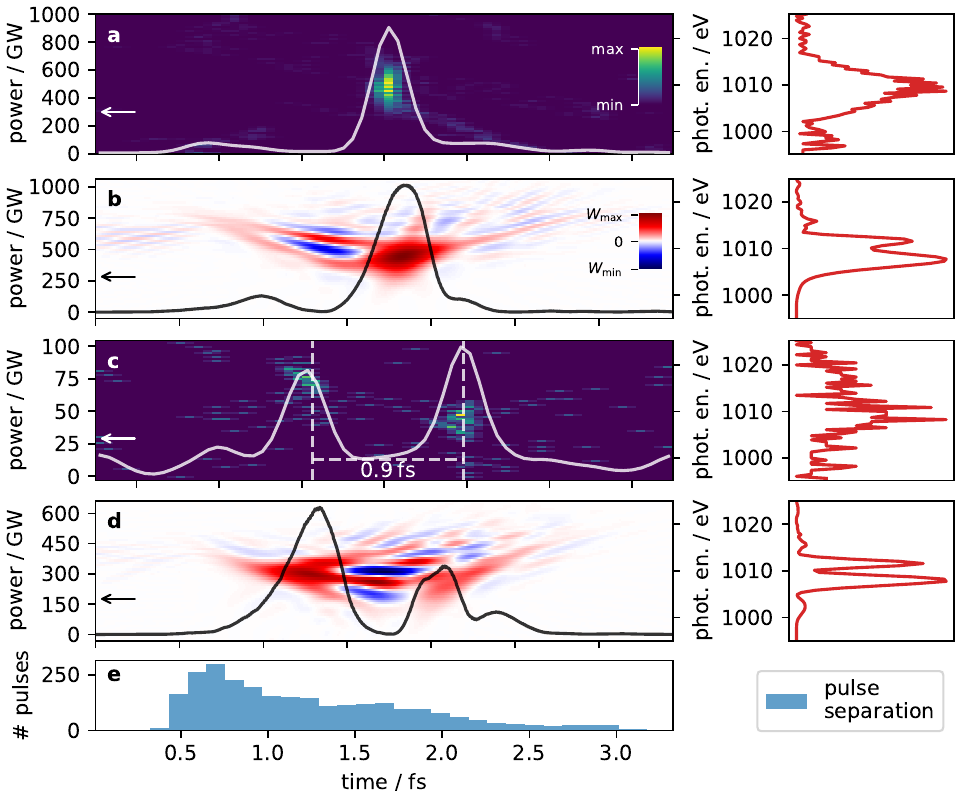}
    \caption{\textbf{a), c)} Examples of reconstructed attosecond FEL pulses. The colour-coded background represents the reconstructed spectrograms (full spectro-temporal information). Overlaid, the temporal shape of the FEL pulses is shown in white. The latter is normalised to the pulse energy to obtain an absolute power figure. The pulse energy is measured independently by the SQS X-ray gas monitor detector (XGMD). The right panels show the spectrograms' projection onto the spectral axis in red, representing an X-ray photon energy spectrum. \textbf{a)} Single-peak 1 terawatt, 250 attosecond FWHM FEL pulse, \textbf{c)} a double-peak pulse.  \textbf{b), d)} Examples of simulated attosecond FEL pulses. In this case, the colour-coded background represents the Wigner distribution of the simulated electric field. The slightly larger energy bandwidth in the experimental case can be attributed to the instrument response, as detailed in the SI. \textbf{e)} Histogram of temporal pulse separation within the dataset of measured double pulses.}
    
    \label{fig:IAS}
\end{figure}

\autoref{fig:IAS} \textbf{a)–c)} illustrate exemplary pulse structures in this regard, with \textbf{a)} depicting a single-peak terawatt attosecond pulse, while \textbf{b)} and \textbf{c)} show double-peak structures with variable delay. These pulses have the required intensities to ultimately enable time-resolved insights into ultrafast \emph{nonlinear} electronic processes as described in the next section. For comparison, in \autoref{fig:IAS} \textbf{e)}, the result of a full SASE FEL pulse simulation based on the actual experimental accelerator parameters is shown. The pulse shape and spectrum derived from the simulated Wigner distribution are in very good qualitative agreement with the general X-ray pulse characteristics from the corresponding measurements in panels \textbf{b)} and \textbf{c)}. In panel \textbf{d)} we depict the distribution of the measured temporal pulse
separations for the subset of double pulses within our data. One can see that it is possible to map out a delay scan spanning 0.5\,fs to 2\,fs in 100\,as steps, just by making use of the inherent SASE pulse variability and sorting on the determined FEL pulse characteristics.

\section*{Attosecond spectroscopy of transient states in neon}
\label{sec:NonlinearMatter}

Due to the high intensity of the XFEL pulses, ionic or excited states of matter can be created efficiently \cite{sorokin2007photoelectric, young2010femtosecond, rudek2012ultra, rudenko2017femtosecond, mazza2020mapping, boll2022x}, and the emerging dynamics can be interrogated via the very same pulse with fully determined characteristics. We show this capability by simultaneously characterising attosecond X-ray pulse shapes and the duration-dependent formation of double-core-hole states in gaseous neon.

For this, we used the adaptability of our detector array and the flexibility of the iterative retrieval code for tuning separate TOF detectors in the spectrometer setup to two different observation regions with respect to well-resolved electron energies. This measurement scheme is also depicted in \autoref{fig:exp}: Twelve detectors are kept at the previously mentioned electron energy at around 120\,eV, given by direct photoionisation from the neon 1s shell by a 990\,eV X-ray pulse, which is used for the pulse reconstruction. Four TOF detectors are tuned to resolve much higher kinetic electron energies, mainly between 760\,eV and 920\,eV, stemming from various Auger processes in the sample after exposure to intense X-ray pulses.

For a broad variety of relatively light elements with atomic numbers $Z<30$, the Auger decay after K-shell ionisation or excitation is the predominant relaxation mechanism, occurring typically within a few femtoseconds.
Studying core-ionised systems \emph{before} the energy is transferred to the secondary Auger electron, promises insights into the localised electronically excited state, preceding and thus governing any kind of X-ray-induced dynamics including subsequent structural changes in more complex molecules.

As shown in earlier studies of gaseous atomic neon, highly intense pulses of XFELs also allow for sequential double photoionisation, where a core-ionised state is created by a first photon and this short-lived intermediate state is further core-ionised or core-excited by a second photon \cite{young2010femtosecond}. In the photon picture, for such a process to occur, it is necessary that the absorption of the second photon precede the Auger decay of the single-core-hole state, which has a lifetime of 2.4\,fs \cite{haynes2021clocking}. This second photon's energy needs to exceed a significantly increased binding energy for the remaining K-shell electron in the transient neon 1s$^{-1}$ system, which amounts to $\sim$990\,eV \cite{mazza2020mapping}. In this case, the core shell of the atom is left empty, and the overall vacancy is termed a double-core hole (DCH). Even more complex dynamics emerge via the multitude of available DCH Auger decay channels if the second K-shell electron is not promoted to the continuum, but excited into one of the Rydberg states of the transient system~\cite{mazza2024resonant}.

\onecolumngrid

\begin{figure}
    \includegraphics[width=0.65\linewidth]{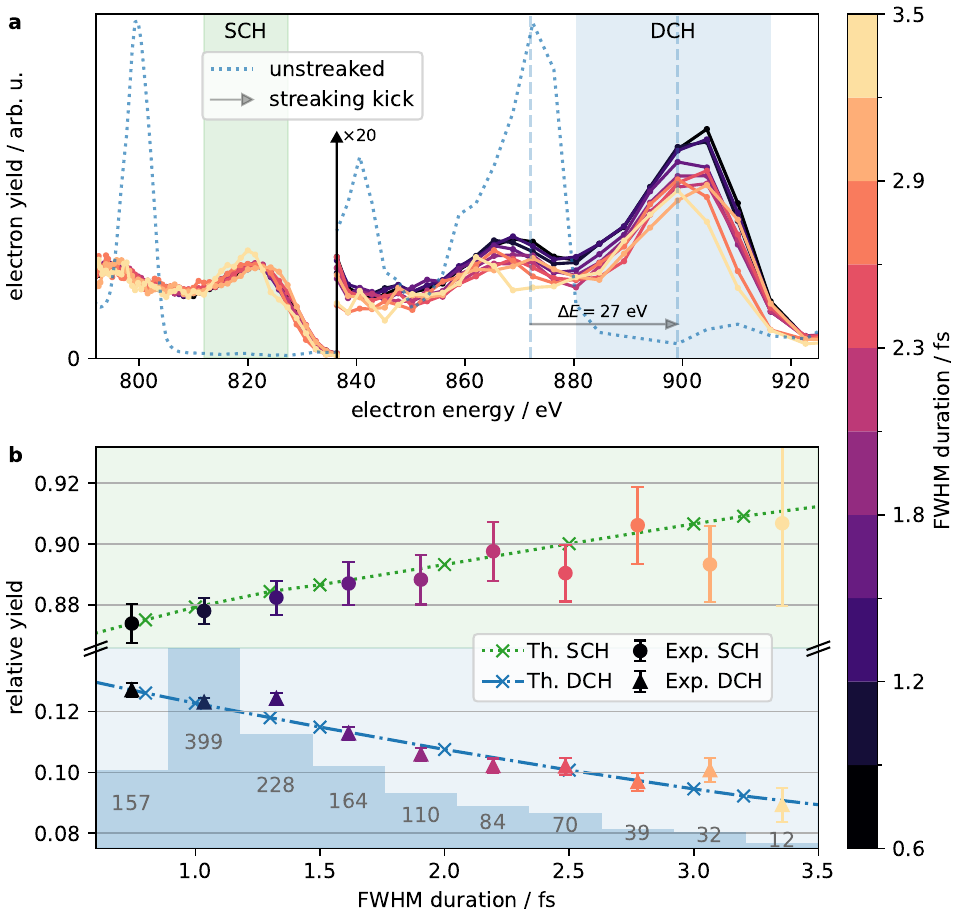}
    \caption{\textbf{a)} Electron spectra of neon irradiated with intense 990\,eV FEL pulses. The colour-coded lines show electron spectra with the MIR laser on (“streaked”), filtered for a specific streaking laser phase and sorted by pulse duration. Furthermore, only shots within a narrow pulse energy window of 200\,$\pm$\,30\,µJ are selected and averaged. For comparison, an averaged spectrum with the MIR laser off (“unstreaked”, light-blue dotted line) is also depicted. The $KL_{2,3}L_{2,3}$ single-core-hole Auger line (SCH) is visible at an energy around 800\,eV, followed by resonant and non-resonant double-core-hole Auger peaks (DCH) between 840\,eV and 880\,eV (signal enlarged 20 times for better visibility). The grey arrow indicates the energy shift induced by the MIR field, when compared to the spectrum without streaking laser.
    \textbf{b)}~The spectrum in \textbf{a)} is integrated over the light-green and light-blue shaded areas to produce a relative yield curve as a function of pulse duration (circles, triangles) for the SCH and DCH peaks. The data clearly reveals that even for a constant pulse energy the fraction of DCH to SCH states increases with shorter pulse durations. A theoretical model for both yields based on rate equations is shown as green dotted (SCH) and blue dash-dotted (DCH) lines. The experimental data points are normalised such that the lowest-duration point in each series lies on the theory curve. The error bars represent the standard error of the sample mean, propagated through integration.  The blue bars with inscribed numbers in the lower half show the number of shots averaged to yield each respective data point.}
    \label{fig:DCHspec}
\end{figure}
\twocolumngrid
In panel \autoref{fig:DCHspec} \textbf{a)} we show electron spectra of different Auger decay channels in neon, from both, single-core-hole (SCH) and double-core-hole states, with the MIR streaking laser turned on (coloured lines) or off (dashed blue line). The coloured lines are measurements of the streaked Auger signals, which were taken simultaneously with the X-ray pulse reconstruction spectra from the streaked neon 1s-electron photolines for the same streaking phase at maximum momentum shift.
The line colour indicates subsequently lower pulse durations of the X-ray pulses, decreasing from yellow to black. The total FEL pulse energy is restricted to a narrow window of 200\,µJ\,$\pm$\,30\,µJ, keeping the total number of photons approximately constant.
Our results demonstrate the clear dependence on individual pulse duration and thus peak-power of the nonlinear DCH signal increase \cite{serkez2018overview,inhester2016xray}, under otherwise unchanged FEL conditions (more details on relative pulse energy stability in the SI, especially Fig. S9).

Investigating this dependence further, we integrate the SCH and DCH peak signals within the green and blue shaded areas, which results in the signal yields plotted as circular and triangular symbols in panel \textbf{b)}. Both experimental data sets are compared to simulation results (crosses; dotted green for SCH, dash-dotted blue for DCH), based on an approach of solving a set of coupled rate equations using calculated atomic cross sections and Auger-decay lifetimes \cite{bhalla1937theoretical, gryzlova2023evolution}. Details about the employed theoretical methods and the rate-equation calculations can be found in the SI.

Several considerations need to be taken into account for the interpretation of the displayed pulse duration dependence of the DCH and SCH signals. Under the given conditions, in the first step for DCH generation, neon atoms are singly core-ionised by the X-ray pulse. Without taking nonlinear processes into account, the amount of SCH Auger electrons will predominantly be determined by the total number of X-ray photons, i.e. the signal from single-photon-absorption processes scales linearly with the pulse energy and is independent of the FEL pulse duration. As mentioned above, for the data shown in \autoref{fig:DCHspec} \textbf{b)} we kept the pulse energy and thus the number of photons practically constant, leaving the number of initially generated SCH states unchanged. Since the next step for creating a DCH requires a second photon \emph{in direct sequence before the Auger relaxation can happen}, reducing the pulse duration, and thus increasing the \emph{peak} power, will in turn increase the DCH yield by enhancing the probability for further ionising the core-ionised neon atoms. That also implies that for every electron detected after promotion of the singly-ionised neon atom to the DCH state, the respective SCH decay channel is closed. Therefore, we expect a counter-acting decline of the SCH electrons in step with the signal increase due to the nonlinear DCH process.

The DCH and SCH signals presented in panel \textbf{b)} show the expected increase of DCH signal with decreasing FWHM \cite{serkez2018overview} of the reconstructed intensity profiles, and thus a direct dependence on the peak power of single X-ray shots. This highlights the ability to employ attosecond-scale knowledge about individual SASE XFEL pulses to uncover essential details of time-dependent nonlinear effects, based on the measurement of shot-to-shot X-ray pulse shape variations within a single FEL operation mode.  These combined measurements not only uncover the nonlinear dependency between X-ray pulse duration and DCH generation probability, but further verify the attosecond X-ray pulse shape reconstruction due to its direct correlation with the DCH dynamics.

In our measurement, we focus on a range of X-ray pulse durations in close vicinity to the expected Auger decay time around 2.4 fs for the SCH state in neon. Locally, the dependence of the DCH signal yield on X-ray duration can be well-fitted by a linear response curve, which is only expected to show its exponential decay characteristics over a broader interval of pulse-duration values (see Fig. S9 in the SI). Higher-order effects due to saturation \cite{Hoener2010, young2010femtosecond} or additional ionisation of valence electrons and subsequent shifts in excitation energies are mostly beyond the scope of the current study and require more sophisticated simulations based on full quantum-mechanical descriptions.
% New section with respect to criticsm from reviewers!
In addition, the chaotic behaviour of SASE electric field statistics can lead to enhanced formation of the DCH state even for long FEL pulses with high enough intensity \cite{PhysRevA.76.033416}.

Nevertheless, the main process investigated in this paper, namely the transition from $\mathrm{Ne}^+$ to dominating $\mathrm{Ne}^{2+}$ states in the sample happens within the SCH Auger decay time, thus within 2.4\,fs in Ne. This is also evident in according time-dependent Boltzmann electron kinetic simulations for high XFEL ($> 10^{18}$\,W/cm²) intensity irradiation of neon, showing a pronounced shift in the full Auger electron energy distribution taking place between 1\,fs and 2\,fs after ionization \cite{Abdallah_2013}.

In our experiment, we used FEL X-ray pulses with pulse durations close to the coherence time of the applied mode of FEL operation, thus consisting of only a single or very few temporal intensity spikes. The findings can thus shed new light on the explicit mechanisms of 1s DCH generation and the dynamical competition between SCH state decay and further excitation in neon. Recent calculations existing down to 15\,fs-XFEL pulses \cite{atoms9040114} could be extended and compared with the results presented here for the even faster dynamical regime opened by attosecond X-ray excitation pulses. Furthermore, comparing with the results of recent simulations for state-resolved ionization dynamics in neon \cite{PhysRevA.107.013102} we experimentally confirm the expected behaviour of the suppression of the $\mathrm{Ne}^{2+}$ DCH decay signal at the highest Auger electron energy for longer X-ray pulses of the order of 100\,fs in comparison with otherwise similar processes triggered by short 1\,fs-pulses.

The same mechanism also leads to so-called frustrated absorption in the neon target when the X-ray pulse duration is short compared to the Auger decay time, as previously demonstrated in molecular nitrogen \cite{PhysRevLett.104.253002} at markedly longer Auger decay times. It is open to further analysis to look deeper into the ultrafast dynamics of processes from different Auger decay channels and to investigate the sequential order of relaxation pathways, which are predicted to happen a bit earlier, when the corresponding probability is higher.
% End of new section.

\section*{Summary and Outlook}

We have reported on the generation of isolated attosecond pulses of up to the TW peak-power regime with durations of only about 200\,as based on the chirp-dispersion scheme at the European XFEL. This approach is straightforward to implement within the existing infrastructure of current SASE FELs and together with the shot-wise temporal pulse characterization opens up the potential for high-repetition rate ultrafast XFEL experiments, such as the exploration of transient states,  high-intensity coherent X-ray diffraction, imaging measurements \cite{thibault2008high, ho2020role}, and other investigations of nonlinear X-ray absorption mechanisms on attosecond time scales.

As a proof of concept, we also demonstrate ML-based prediction of iteratively reconstructed XFEL pulse profiles from experimental detector images (see Sec.~S2.6 in the SI). Our long-term vision is to integrate this approach into the beamline data readout to provide single-shot pulse information at the full XFEL repetition rate, supporting online feedback and ultimately enabling closed-loop experimentation.

Enabled by the high intensity of the employed attosecond pulses at 990\,eV, we have shown new paths to nonlinear spectroscopy of highly transient states of matter with X-rays. The state-specific exploration of nonlinear electron dynamics via coherent attosecond X-ray pulses is a promising perspective towards exploring the physical origins of reaction processes together with the subsequent chemical evolution and ultimately a system's functionality.

While angular streaking allows for studying details of electronic motion and redistribution, it can also readily map out their translation into coupled nuclear dynamics.
This method may be used in the future to directly follow electron rearrangement after excitation of specific elements in complex molecules, with the goal to sense and control the electronic evolution in the origins of chemistry with atomic precision and to understand radiation damage mechanisms on a more fundamental level.
Intriguing prospects of tracking electron migration in larger molecules such as peptides and amino acids and understanding the mechanisms of functionality-altering energy deposition and subsequent molecular restructuring are opened up. Combining the presented modes with the future possibilities of X-ray polarisation control at FELs will even allow studying the time-dependent asymmetric structures of such systems and their chiral dynamics.

%\backmatter

\begin{acknowledgements}
%\bmhead{Acknowledgements}
We thank the European XFEL for provision of beamtime under proposal number \#2828 and \#6176 and the many support groups for their sedulous effort. We also thank the DESY photon-science workshop under the lead of Markus Kowalski for their kind and competent help. We thank Elena V. Gryzlova for providing calculations of photoionisation cross sections for the rate equation model. We furthermore thank Raimund K\"ammering, serving as the run coordinator and providing invaluable support to our activities in the XFEL control room during our experiment. This work has been supported by the Bundesministerium f\"ur Forschung Technologie und Raumfahrt (BMFTR) under grant 13K22CHA. M.I., A.E., A.Ha., L.M. acknowledge support from the Deutsche Forschungsgemeinschaft (DFG)-Project No. 328961117-SFB 1319 ELCH (Extreme light for sensing and driving molecular chirality). L.F., K.S., A.E., A.Ha., A.He., R.H., L.M., D.M., S.S., B.S., L.W., M.I. and W.H.  acknowledge funding of the BMBF-ErUM-Pro project “TRANSALP” (05K22PE3) and the BMBF project “SpeAR\_XFEL” (05K19PE1). M.I. and T.M. acknowledge funding of the BMFTR-ErUM-Pro project “AttoSee” (05K25GU4). M.I., K.S., and B.S. acknowledge funding of the BMFTR-ErUM-Data project “OPTIX” (05D2025). M.I. and V.W. acknowledge funding from the Deutsche Forschungsgemeinschaft - project "CACTUS" (707468). J.P.M. and F.E. acknowledge funding from EPSRC/UKRI EP/X026094/1 and EP/V026690/1 as well as UK Physical Sciences XFEL Hub. This work is supported by the Cluster of Excellence 'CUI: Advanced Imaging of Matter' of the Deutsche Forschungsgemeinschaft (DFG) – EXC 2056 – project ID 390715994. P.R. acknowledges funding from the Alexander von Humboldt Foundation (Feodor Lynen Fellowship). M.L., and V.Z. thank the support from the Swedish Research Council (VR) through the Röntgen-Ångström Cluster program (grant No. 2021-05967). S.B. acknowledges funding from the Helmholtz Initiative and Networking Fund. P.R. and T.P. acknowledge support by the Deutsche Forschungsgemeinschaft (DFG, German Research Foundation) under Germany’s Excellence Strategy EXC2181/1-390900948 (the Heidelberg STRUCTURES Excellence Cluster). Data recorded for the experiment at the European XFEL are available at \doi{10.22003/XFEL.EU-DATA-002828-00} and \doi{10.22003/XFEL.EU-DATA-006176-00}.
\end{acknowledgements}

\def\bibcommenthead{}%

\bibliography{refs_new}

\end{document}